\documentclass[12pt]{article}
\usepackage{graphicx}

\newcommand{\munu}{^\mu_{\phantom{\mu}\nu}}
\makeatletter
\@addtoreset{equation}{section}

\makeatother

\begin{document}
\begin{center}
{\large{\bf  Relativistic Einstein-Podolsky-Rosen correlation
and Bell's inequality}}
\vskip .5 cm
{\bf Hiroaki Terashima and Masahito Ueda}
\vskip .4 cm
{\it Department of Physics, Tokyo Institute of Technology\\
Tokyo 152-8551, Japan}
\vskip 0.5 cm
\end{center}

\begin{abstract}
We formulate the Einstein-Podolsky-Rosen (EPR)
gedankenexperiment within
the framework of relativistic quantum theory
to analyze a situation in which measurements
are performed by moving observers.
We point out that under certain conditions
the perfect anti-correlation of an EPR pair of spins
in the \emph{same} direction
is deteriorated in the moving observers' frame
due to the Wigner rotation,
and show that the degree of the violation
of Bell's inequality \emph{prima facie}
decreases with increasing
the velocity of the observers
if the directions of the measurement are fixed.
However, this does not imply
a breakdown of non-local correlation
since the perfect anti-correlation is maintained in
appropriately chosen \emph{different} directions.
We must take account of this relativistic effect
in utilizing in moving frames
the EPR correlation and the violation of Bell's inequality
for quantum communication.
\end{abstract}

\begin{flushleft}
\end{flushleft}

\section{Introduction}
In the context of physical reality,
Einstein, Podolsky, and Rosen (EPR)~\cite{EiPoRo35}
proposed a gedankenexperiment.
Bohm's version~\cite{Bohm89} of this experiment
is described as follows:
Suppose that two spin-$1/2$ particles
with total spin zero are moving in opposite directions
and that their spin components are measured by two observers.
The spin component of each particle is
measured in some direction by each observer.
If the two measurements are performed in the same direction,
the outcomes are perfectly anti-correlated
in whichever direction and at however remote place
they are performed.
This remarkable property is known as the EPR correlation.
The reason for the perfect anti-correlation
in any direction is that the spin-singlet state is isotropic and
has no preferred direction.

In this paper, we consider Bohm's version of
the EPR experiment in the framework of
relativistic quantum theory and discuss an effect
of observer's motion on the EPR correlation.
A number of articles~\cite{Bloch67,HelKra70,AhaAlb,%
HelKon83,GhGrPe90,SuaSca,ZBGT01}
have discussed the case of a moving observer
in connection with the instantaneous reduction of
the state caused by the measurement
in relativistic quantum theory.
However, we shall not consider such a problem
and concentrate on the symmetry transformation of
the spin correlation under the Lorentz transformation.

To address the issue of how the spin
of a particle is seen from a moving observer,
we need not resort to the explicit dynamics of the particle
such as the Dirac equation,
but only to a group-theoretical approach~\cite{Ohnuki88,Weinbe95}.
This is analogous to the fact that
the transformation law of the spin
under the spatial rotation
is derived only from knowledge of
the spatial rotation group $SO(3)$.
In relativistic theory, the relevant group is
the Poincar\'{e} group $ISO(1,3)$
which contains $SO(3)$ as a subgroup.
For a massive particle (such as an electron)
and a massless particle (such as a photon),
transformation laws under the Lorentz transformation
are derived from the unitary representations of
the Poincar\'{e} group.
Note that massless particles are entirely different
from massive particles in relativistic theory
since massless particles
move at the speed of light and thus have no rest frame.
In the massive case,
the Lorentz transformation rotates the spin of a particle,
known as the Wigner rotation~\cite{Wigner39}.
The angle of this rotation depends
not only on the Lorentz transformation
but also on the momentum of the particle.
On the other hand, in the massless case,
the Lorentz transformation rotates
the plane of polarization.

Recently, the Wigner rotation has seen a remarkable resurgence
of interest in the context of
the entanglement or the EPR correlation in the relativistic
regime~\cite{PeScTe02,AlsMil02,TerUed02,GinAda02,AhLeHw}.
Peres, Scudo, and Terno~\cite{PeScTe02} have shown that
the spin entropy for a single particle is not Lorentz invariant
if the particle is not in the momentum eigenstate,
because spin is entangled with momentum by the Wigner rotation.
Gingrich and Adami~\cite{GinAda02} have shown that
entanglement between the spins of two particles is carried over
to the entanglement between the momenta of the particles
by the Wigner rotation,
even though the entanglement of the entire system
is Lorentz invariant.
Alsing and Milburn~\cite{AlsMil02} have
considered entanglement in one of the Bell states
between the spins of two particles 
moving in opposite directions with definite momenta,
and shown that it is Lorentz invariant
since the Wigner rotation is a local unitary operation.
Instead of the state vector in the Hilbert space,
they have used a 4-component Dirac spinor or
a polarization vector in favor of quantum field theory.
Independently of their research,
the authors~\cite{TerUed02} considered
a similar situation but discussed the EPR correlation
rather than the entanglement using the spin-singlet state
in terms of the state vector.
(The entanglement is independent of the basis
for the measurement, but the correlation depends on it.)
Later, Ahn et al.~\cite{AhLeHw} also
calculated the same situation
with all the Bell states and obtained
a conclusion on the EPR correlation that contradicts our result.
They concluded that the Wigner rotation could cause
``a counter example for the nonlocality of the EPR paradox''.

This paper is an extended
and comprehensive version of Ref.~\cite{TerUed02},
discussing a possible experiment and
relevance to quantum communications.
We first consider two massive particles
in the spin-singlet state
moving in opposite directions with definite momenta,
and two observers moving in the same direction
at the same velocity with respect to the laboratory frame.
By applying the Wigner rotation,
we show that
the measurement results are not perfectly anti-correlated
when the two observers measure the spins
in the same direction in their common rest frame.
Seen from the moving observers,
the anti-correlation in the same direction decreases
unlike the non-relativistic case.
In particular, the perfect anti-correlation
in every direction is not maintained in all inertial frames.
The special directions along which the perfect anti-correlation
is not maintained are specified
by the motion of the observers and by that of the particles.
We also examine the violation of
Bell's inequality~\cite{Bell64,CHSH69} in this case
and show that the motion of the observers
\emph{prima facie} decreases the degree of violation
of Bell's inequality if the directions of the measurement are fixed.
We extend these considerations to the massless case
and obtain qualitatively similar but quantitatively different results.

At first sight,
the above results appear to contradict
the nonlocality inherent in quantum mechanics.
One might think that
the quantum correlation breaks down and
that local hidden variable theories are restored
in the relativistic regime.
However, this is not the case, of course.
We explicitly show that the perfect anti-correlation
is maintained in \emph{different} directions and
Bell's inequality for an appropriate set of observables
is still maximally violated.
Note that the entanglement is preserved
under the Lorentz transformation~\cite{AlsMil02}.
We emphasize that
our aim is not to discuss the foundation of quantum mechanics
but to explore effects of the relative motion
between the sender and receiver in quantum communications
that utilize the EPR correlation and
the violation of Bell's inequality.

This paper is organized as follows:
In order to make this paper self-contained,
Sections~\ref{sec:massive} and \ref{sec:massless}
review the Lorentz transformation laws
for a massive particle and for a massless particle,
respectively.
Section~\ref{sec:epr} formulates relativistic EPR experiments
both for massive particles and for massless particles,
and applies the transformation laws to them.
Section~\ref{sec:bell} analyzes
Bell's inequality
in the relativistic EPR experiments.
Section~\ref{sec:discuss} summarizes our results, and
discusses an experimental possibility and
relevance to quantum information.
Appendix summarizes the Poincar\'{e} group
and our conventions.

\section{\label{sec:massive}Massive Particle}
In this section,
we follow Ref.~\cite{Weinbe95} to discuss
the one particle states for a massive particle
which furnish an irreducible representation
of the Poincar\'{e} group.
We then calculate the transformation law explicitly
in a specific example.

First, we consider the rest frame of
the massive particle,
since the spin is most conveniently
identified in the rest frame.
The four-momentum of the particle with mass $M$
then becomes the rest momentum,
\begin{equation}
 k^\mu=(Mc,0,0,0).
\label{stdk}
\end{equation}
In this frame,
the state $|k,\sigma\rangle^{\textrm{rest}}$ is specified
in terms of the eigenvalues of the Hamiltonian $H$,
the momentum operator $\vec{P}$, and
the $z$-component of
the total angular momentum operator $\vec{J}$ as
\begin{eqnarray}
H |k,\sigma\rangle^{\textrm{rest}}
     &=& Mc^2 |k,\sigma\rangle^{\textrm{rest}}, \\
\vec{P} |k,\sigma\rangle^{\textrm{rest}} &=& 0, \\
J^3 |k,\sigma\rangle^{\textrm{rest}}   
    &=& \sigma\hbar |k,\sigma\rangle^{\textrm{rest}}.
\end{eqnarray}
For a spatial rotation $R\munu$,
there is a corresponding unitary operator $U(R)$
which can be represented by
a $(2j+1)$-dimensional unitary matrix $D^{(j)}(R)$,
\begin{equation}
U(R)\, |k,\sigma\rangle^{\textrm{rest}}  = \sum_{\sigma'}
            D_{\sigma'\sigma}^{(j)}(R)\,|k,\sigma'\rangle^{\textrm{rest}},
\label{rotj}
\end{equation}
where $j$ is an integer or a half-integer
and $-j\le\sigma\le j$.
Note that the momentum on the right-hand side is also
the rest momentum $k^\mu$ since
the spatial rotation group $SO(3)$
leaves $k^\mu$ invariant.
That is, the particle remains at rest
if the frame is rotated spatially.
The total angular momentum
is equal to the spin in the rest frame
because the orbital angular momentum is absent.
Therefore, $j$ is the spin of the particle and
$\sigma$ its $z$-component.

We next consider the general frame,
where the four-momentum of the particle is
\begin{equation}
 p^\mu=(\sqrt{|\vec{p}|^2+M^2c^2},p^1,p^2,p^3).
\end{equation}
We obtain this momentum by performing
a standard Lorentz transformation $L(p)\munu$
on the rest momentum (\ref{stdk}),
\begin{equation}
   p^\mu=L(p)\munu k^{\nu}.
\label{deflp}
\end{equation}
An explicit form of $L(p)\munu$ is written as
\begin{eqnarray}
L(p)^0_{\phantom{0}0} &=& \gamma, \nonumber \\
L(p)^0_{\phantom{0}i} &=& L(p)^i_{\phantom{i}0}=p^i/Mc,    \\
L(p)^i_{\phantom{i}k} &=& \delta_{ik}+
(\gamma-1)\,p^i\,p^k/|\vec{p}|^2,\nonumber
\end{eqnarray}
where
\begin{equation}
\gamma=\frac{\sqrt{|\vec{p}|^2+M^2c^2}}{Mc}.
\end{equation}
Note that the inverse matrix
$L^{-1}(p)\munu$ of $L(p)\munu$ can be obtained
by replacing $p^i$ by $-p^i$,
i.e., $L^{-1}(p^0,p^i)\munu=L(p^0,-p^i)\munu$.

Using the unitary operator $U(L(p))$
corresponding to $L(p)\munu$,
we define the state in the general frame as
\begin{equation}
 |p,\sigma\rangle\equiv\,U(L(p))\, |k,\sigma\rangle^{\textrm{rest}}.
\label{defp}
\end{equation}
Note that
\begin{eqnarray}
H |p,\sigma\rangle
  &=&\sqrt{|\vec{p}|^2c^2+M^2c^4}\,|p,\sigma\rangle, \\
\vec{P}|p,\sigma\rangle
  &=& \vec{p}\,|p,\sigma\rangle,
\end{eqnarray}
where $\vec{p}=(p^1,p^2,p^3)$, but that
\begin{equation}
J^3 |p,\sigma\rangle
  \neq\sigma\hbar\,|p,\sigma\rangle.
\end{equation}
This is because $J^3$ now contains a contribution
from the orbital motion.
Since we have
\begin{eqnarray*}
  \left[J^3, P^1\right] &=& i\hbar P^2\neq 0,  \\
  \left[J^3, P^2\right] &=& -i\hbar P^1\neq 0,
\end{eqnarray*}
$|p,\sigma\rangle$ cannot be a
simultaneous eigenstate of $J^3$ and $\vec{P}$.
Unlike in the rest frame,
the label $\sigma$ denotes not
the eigenvalue of $J^3$ but the $z$-component of the spin.

The 3-dimensional delta function $\delta^3(\vec{p'}-\vec{p})$
is not Lorentz invariant.
Instead,
\begin{equation}
\sqrt{|\vec{p}|^2+M^2c^2}\;\delta^3(\vec{p'}-\vec{p})
\label{invdel}
\end{equation}
is an invariant delta function
with the mass shell condition,
\begin{equation}
p^0=\sqrt{|\vec{p}|^2+M^2c^2}.
\end{equation}
The normalization of the state is thus chosen to be
\begin{equation}
\langle p',\sigma'|p,\sigma\rangle
  =\left(\frac{\sqrt{|\vec{p}|^2+M^2c^2}}{Mc}\right)
  \,\delta^3(\vec{p'}-\vec{p})\,\delta_{\sigma'\sigma}.
\end{equation}

\begin{figure}
\begin{center}
\includegraphics[scale=0.55]{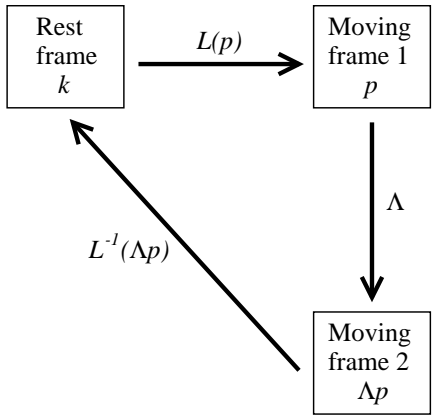}
\end{center}
\caption{\label{fig1}The Wigner rotation.}
\end{figure}

The state defined in Eq.~(\ref{defp})
transforms under the Lorentz transformation
$\Lambda\munu$ as
\begin{eqnarray}
U(\Lambda)\, |p,\sigma\rangle
  &=& U(\Lambda)\,U(L(p))\, |k,\sigma\rangle^{\textrm{rest}} \nonumber \\
  &=& U(L(\Lambda p))\,U(W(\Lambda,p))\, |k,\sigma\rangle^{\textrm{rest}},
\end{eqnarray}
where
\begin{equation}
W(\Lambda,p)\munu=
\left[L^{-1}(\Lambda p)\,\Lambda\,L(p)\right]\munu
\label{defw}
\end{equation}
is called Wigner rotation.
It follows from Eq.~(\ref{deflp})
that the Wigner rotation leaves
the rest four-momentum (\ref{stdk}) invariant,
\begin{equation}
 W(\Lambda,p)\munu k^{\nu}=k^\mu.
\end{equation}
This means that
the Wigner rotation is an element of
the spatial rotation group $SO(3)$.
We can view this Wigner rotation as follows:
we perform the Lorentz transformation $L(p)\munu$
on the rest frame to obtain a moving frame 1,
and then operate the Lorentz transformation $\Lambda\munu$
to obtain another moving frame 2,
as depicted by Fig.~\ref{fig1}.
We return to the rest frame by further performing
the Lorentz transformation 
$L^{-1}(\Lambda p)\munu$ on the moving frame 2.
However, the resultant rest frame is different from
the original one by the Wigner rotation as indicated
by Eq.~(\ref{defw}).

Therefore,
if the spin of the particle is $j$ as in Eq.~(\ref{rotj}),
we find that
\begin{equation}
 U(\Lambda)\, |p,\sigma\rangle
  = \sum_{\sigma'} D_{\sigma'\sigma}^{(j)}(W(\Lambda,p))
    \,|\Lambda p,\sigma'\rangle.
\label{trans}
\end{equation}
Note that the spin part is transformed in a manner
depending on the momentum $p^\mu$, in general.
This dependence plays a crucial role
in later discussions on the EPR correlation.
On the other hand,
the transformation of the spin part
under the spatial rotation is independent of $p^\mu$.
In fact,
when $\Lambda\munu$ is a spatial rotation $R\munu$,
the Wigner rotation $W(R,p)\munu$ reduces to
$R\munu$ itself for all $p^\mu$~\cite{Weinbe95}.
That is, under spatial rotations,
the state $|p,\sigma\rangle$ transforms
in the same way as in non-relativistic quantum mechanics,
\begin{equation}
 U(R)\, |p,\sigma\rangle
  = \sum_{\sigma'} D_{\sigma'\sigma}^{(j)}(R)
    \,|R p,\sigma'\rangle.
\end{equation}

Let us find the transformation law (\ref{trans})
in a specific example.
Suppose that in a laboratory frame the massive particle
is moving along the $x$-axis with four-momentum
\begin{equation}
p^\mu=(Mc\,\cosh\xi,Mc\,\sinh\xi,0,0),
\end{equation}
where the rapidity $\xi$ is related to
the velocity of the particle $v=dx/dt$ by
\begin{equation}
\frac{v}{c}=\tanh\xi.
\end{equation}
The standard Lorentz transformation (\ref{deflp}) becomes
\begin{equation}
L(p)\munu=B_x(\xi)\munu,
\end{equation}
where $B_x(\xi)\munu$ is defined by
\begin{equation}
B_x(\xi)\munu=
  \left(\begin{array}{cccc}
    \cosh\xi &  \sinh\xi & 0 & 0 \\
    \sinh\xi &  \cosh\xi & 0 & 0 \\
           0 &         0 & 1 & 0 \\
           0 &         0 & 0 & 1
  \end{array}\right).
\end{equation}
We then introduce an observer whose velocity $V$ is
given in the laboratory frame by
\begin{equation}
\frac{V}{c}=\tanh\chi.
\label{velobs}
\end{equation}
From the rotational symmetry about the $x$-axis,
it suffices to assume that the observer
is moving in the direction
$(\cos\theta,0,\sin\theta)$ on the $xz$-plane.
The rest frame of the observer is obtained by
performing a Lorentz transformation
\begin{equation}
\Lambda\munu=
\left[R_y(\theta)B_x^{-1}(\chi)
R_y^{-1}(\theta)\right]\munu
\end{equation}
on the laboratory frame.
(Note that the rotation $R_y^{-1}(\theta)\munu$ brings
the direction $(\cos\theta,0,\sin\theta)$ into the $x$-axis.)
In this frame,
the observer describes the state $|p,\sigma\rangle$
as $U(\Lambda)|p,\sigma\rangle$.

A straightforward calculation shows that
the Wigner rotation (\ref{defw}) reduces to a
rotation about the $y$-axis,
\begin{equation}
W(\Lambda,p)\munu=R_y^{-1}(\delta)\munu,
\end{equation}
where
\begin{eqnarray}
\cos\delta &=& \frac{A-B\cos\theta+C\cos^2\theta}%
{D-B\cos\theta}, \\
\sin\delta &=& \frac{B\sin\theta-C\sin\theta\cos\theta}%
{D-B\cos\theta},
\end{eqnarray}
with
\begin{eqnarray}
A &=&  \cosh\xi+\cosh\chi, \\
B &=& \sinh\xi\sinh\chi, \\
C &=& (\cosh\xi-1)(\cosh\chi-1), \\
D &=&  \cosh\xi\cosh\chi+1 .
\end{eqnarray}
The transformation law (\ref{trans}) thus becomes
\begin{equation}
 U(\Lambda)\, |p,\sigma\rangle
  = \sum_{\sigma'} D_{\sigma'\sigma}^{(j)}(R_y^{-1}(\delta))
    \,|\Lambda p,\sigma'\rangle.
\end{equation}
This means that the spin is rotated about the $y$-axis
through the angle $\delta$
in the observer's rest frame.
In general,
it is rotated within the plane spanned by
the motion of the particle and
that of the observer.

For the case of the spin-$1/2$ particle, we have
\begin{eqnarray}
  D_{\sigma'\sigma}^{(1/2)}(R_y^{-1}(\delta)) &=&
    \exp\left(-i\frac{\sigma_y}{2}\delta\right) \nonumber \\
  &=& \left(\begin{array}{cccc}
    \cos\frac{\delta}{2} &-\sin\frac{\delta}{2}   \\[5pt]
    \sin\frac{\delta}{2} & \cos\frac{\delta}{2}
   \end{array}\right),
\end{eqnarray}
where $\sigma_y$ is the Pauli matrix.
The transformation law for the spin-$1/2$ particle is thus given by
\begin{eqnarray}
U(\Lambda)\,|p,\uparrow\rangle &=&
 \cos\frac{\delta}{2}\, |\Lambda p,\uparrow\rangle
 +\sin\frac{\delta}{2}\, |\Lambda p,\downarrow\rangle,  \label{mass1} \\
U(\Lambda)\,|p,\downarrow\rangle &=&
 -\sin\frac{\delta}{2}\, |\Lambda p,\uparrow\rangle
 +\cos\frac{\delta}{2}\, |\Lambda p,\downarrow\rangle,  \label{mass2}
\end{eqnarray}
where $\uparrow=+1/2$ and $\downarrow=-1/2$.

\begin{figure}
\begin{center}
\includegraphics[scale=0.55]{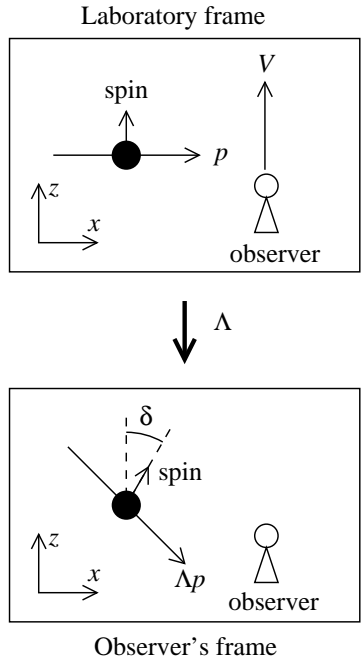}
\end{center}
\caption{\label{fig2}The orthogonal observer.}
\end{figure}

\begin{figure}
\begin{center}
\includegraphics[scale=0.65]{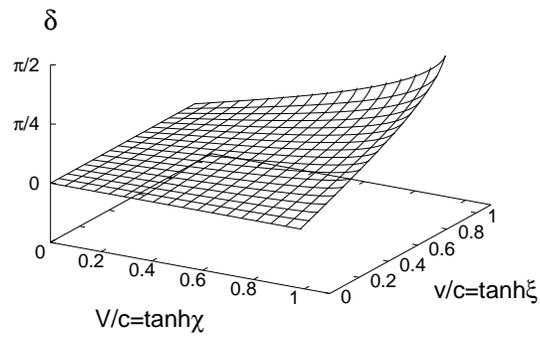}
\end{center}
\caption{\label{fig3}The Wigner angle $\delta$ for the orthogonal
observer as a function of $V/c=\tanh\chi$ and $v/c=\tanh\xi$.}
\end{figure}

If the observer is moving in the same direction
as the particle (i.e. $\theta=0$),
we find that the Wigner angle $\delta$ is zero, and
therefore, the spin part is unchanged by this transformation.
From now on,
we concentrate on a situation in which
the observer is moving in the $z$ direction
(i.e. $\theta=\pi/2$),
as illustrated in Fig.~\ref{fig2}.
In this case,
the Wigner angle $\delta$ is given by (see Fig.~\ref{fig3})
\begin{equation}
\tan\delta = \frac{\sinh\xi\sinh\chi}{\cosh\xi+\cosh\chi}.
\label{defdel}
\end{equation}
When either $\xi=0$ or $\chi=0$,
we find that $\delta$ is $0$.
That is,
when either the particle or the observer is at rest,
the direction of the spin is unchanged.
In the limit of $\xi\to0$ and $\chi\to0$,
the Wigner angle $\delta$ becomes $\xi\chi/2$.
On the other hand,
in the opposite case where $\xi\to\infty$
and $\chi\to\infty$,
the Wigner angle $\delta$ becomes $\pi/2$.
This is the maximal effect, where the spin,
which in the laboratory frame
points in the $z$ direction,
tilts into the $x$ direction in the observer's rest frame.

A physical picture of the Wigner rotation is as follows:
The Lorentz transformation $\Lambda\munu$,
from the laboratory frame to the observer's frame,
rotates the direction of the momentum into
\begin{equation}
\vec{\Lambda p}=(Mc\,\sinh\xi,0,-Mc\,\cosh\xi\sinh\chi),
\end{equation}
as depicted in Fig.~\ref{fig2},
where the angle of this rotation, $\delta_p$, is given by
\begin{equation}
  \tan\delta_p=
  \left|\frac{(\Lambda p)^3}{(\Lambda p)^1}\right|
  =\frac{\sinh\chi}{\tanh\xi}.
\label{momang}
\end{equation}
The spin is dragged by this rotation
since the spin is coupled to the momentum
in relativistic quantum theory.
Note that the angle of the spin rotation is
less than or equal to that of the momentum rotation,
$\delta\le\delta_p$.
In non-relativistic quantum theory,
the Galilean transformation rotates the momentum
by the angle $\tan\delta_p=V/v$ but
does not rotate the spin
since the spin is not coupled to the momentum.
In fact,
it is easy to see that $\delta\to0$
in the non-relativistic limit $c\to\infty$
since this limit is equivalent to
the limit $\xi\to0$ and $\chi\to0$.

We can also understand the Wigner rotation
in terms of the unitary operator.
It is the difference between
$\Lambda L(p)\munu$ and $L(\Lambda p)\munu$ that
gives rise to the Wigner rotation (\ref{defw}),
as illustrated in Fig.~\ref{fig1}.
These are not equal,
even though both of them
bring the momentum $k^\mu$ to $\Lambda p^\mu$.
In the limit of $\xi\to0$ and $\chi\to0$,
\begin{eqnarray}
U(\Lambda)U(L(p)) &=&
  e^{-\frac{i}{\hbar}K^3\chi}
  e^{\frac{i}{\hbar}K^1\xi}, \\
U(L(\Lambda p)) &\simeq&
 e^{\frac{i}{\hbar}
     \left(-K^3\chi+K^1\xi\right)},
\end{eqnarray}
where $\vec{K}$ is the boost operator (\ref{defK}).
Using the Baker-Campbell-Hausdorff formula and
the commutation relation $\left[K^1, K^3\right]=i\hbar J^2$,
we obtain
\begin{equation}
U(\Lambda)U(L(p))\simeq U(L(\Lambda p))\,
   e^{-\frac{i}{\hbar}J^2 \frac{\xi\chi}{2}}.
\end{equation}
This means that
$U(\Lambda)\,|p,\sigma\rangle$ is
different from $|\Lambda p,\sigma\rangle$
by the rotation of the spin about the $y$-axis, because
\begin{eqnarray}
U(\Lambda)\,|p,\sigma\rangle &=&
 U(\Lambda)U(L(p))\,|k,\sigma\rangle^{\textrm{rest}} \nonumber \\
      &\simeq& U(L(\Lambda p))\,
   e^{-\frac{i}{\hbar}J^2 \frac{\xi\chi}{2}} 
  \,|k,\sigma\rangle^{\textrm{rest}},
\end{eqnarray}
but
\begin{equation}
|\Lambda p,\sigma\rangle =
    U(L(\Lambda p))\, |k,\sigma\rangle^{\textrm{rest}}.
\end{equation}

\section{\label{sec:massless}Massless Particle}
We again follow Ref.~\cite{Weinbe95} to
discuss the one particle states for a massless particle
and calculate the transformation law in a specific example.

The main difference from the massive case is that
massless particles cannot be at rest
in any reference frame.
It is convenient to consider a standard frame
in which the four-momentum becomes a standard momentum
\begin{equation}
 k^\mu=(\kappa,0,0,\kappa),
\label{stdk0}
\end{equation}
with $\kappa$ $(>0)$ held fixed.
In this frame,
the state $|k,\sigma\rangle^{\textrm{std}}$ is specified
in terms of the eigenvalues of the Hamiltonian $H$,
the momentum operator $\vec{P}$, and
the $z$-component of
the total angular momentum operator $\vec{J}$ as
\begin{eqnarray}
H |k,\sigma\rangle^{\textrm{std}}
       &=& \kappa c|k,\sigma\rangle^{\textrm{std}}, \\
P^1 |k,\sigma\rangle^{\textrm{std}}
       &=& P^2 |k,\sigma\rangle^{\textrm{std}}=0,  \\
P^3 |k,\sigma\rangle^{\textrm{std}}
       &=& \kappa |k,\sigma\rangle^{\textrm{std}}, \\
J^3 |k,\sigma\rangle^{\textrm{std}}
       &=& \sigma\hbar |k,\sigma\rangle^{\textrm{std}}.
\end{eqnarray}
Note that $\sigma$ means the helicity, that is,
the component of angular momentum
in the direction of motion.

The symmetry of this system is no longer
the spatial rotation group $SO(3)$
since the particle is moving in the $z$ direction.
The symmetry of the standard momentum $k^\mu$ is $ISO(2)$,
whose general element $T\munu$ can be decomposed into
\begin{equation}
T(\alpha,\beta,\gamma)\munu=
\left[S(\alpha,\beta)R_z(\gamma)\right]\munu,
\label{iso2}
\end{equation}
where $R_z(\gamma)\munu$ is
the rotation about the $z$-axis
through the angle $\gamma$ and
\begin{equation}
S(\alpha,\beta)\munu
  =\left(\begin{array}{cccc}
    1+\zeta & \alpha & \beta & -\zeta   \\
    \alpha  &      1 &     0 & -\alpha  \\
    \beta   &      0 &     1 & -\beta   \\
    \zeta   & \alpha & \beta & 1-\zeta
   \end{array}\right), \quad
\end{equation}
with $\zeta=(\alpha^2+\beta^2)/2$ being
the ``translation'' by
the vector $(\alpha,\beta)$ in the sense that
\begin{equation}
 \left[S(\bar{\alpha},\bar{\beta})
    S(\alpha,\beta)\right]\munu
  =S(\bar{\alpha}+\alpha, \bar{\beta}+\beta)\munu.
\end{equation}
The unitary operator corresponding to
$T\munu$ is given by
\begin{eqnarray}
U(T(\alpha,\beta,\gamma)) &=&
    U(S(\alpha,\beta))U(R_z(\gamma))\nonumber \\
 &=& e^{i\alpha A+i\beta B}e^{iJ^3\gamma},
\end{eqnarray}
where $A=J^2+K^1$ and $B=-J^1+K^2$.
However, in the case of the photon,
the eigenvalues of $A$ and $B$ are both zero.
We thus have
\begin{equation}
U(T(\alpha,\beta,\gamma))\, |k,\sigma\rangle^{\textrm{std}}
 = e^{i\gamma\sigma}\,|k,\sigma\rangle^{\textrm{std}},
\label{hel0}
\end{equation}
for the transformation $T\munu$ belonging to $ISO(2)$.

We next consider the general frame,
where the four-momentum of the particle is given by
\begin{equation}
 p^\mu=(|\vec{p}|,p^1,p^2,p^3).
\end{equation}
We obtain this momentum by performing
a standard Lorentz transformation $L(p)\munu$
on the standard momentum (\ref{stdk0}),
\begin{equation}
p^\mu=L(p)\munu k^{\nu},
\label{deflp0}
\end{equation}
where
\begin{equation}
L(p)\munu=
\left[R_z^{-1}(\varphi)R_y^{-1}(\vartheta)
B_z(\ln(|\vec{p}|/\kappa))\right]\munu,
\end{equation}
and
\begin{equation}
\hat{p}=\vec{p}/|\vec{p}|=
(\sin\vartheta\cos\varphi,\sin\vartheta\sin\varphi,\cos\vartheta)
\label{resang}
\end{equation}
with $0\le\vartheta\le\pi$ and $0\le\varphi<2\pi$.
Note that the rotation
$[R_z^{-1}(\varphi)R_y^{-1}(\vartheta)]\munu$
brings the $z$-axis into
the direction of $\hat{p}$.

Using the unitary operator corresponding to $L(p)\munu$,
we define the state in the general frame as
\begin{equation}
 |p,\sigma\rangle\equiv\,U(L(p))\, |k,\sigma\rangle^{\textrm{std}}.
\label{defp0}
\end{equation}
Note again that
\begin{eqnarray}
H |p,\sigma\rangle
  &=&|\vec{p}|c\,|p,\sigma\rangle, \\
\vec{P}|p,\sigma\rangle
  &=& \vec{p}\,|p,\sigma\rangle,
\end{eqnarray}
but that
\begin{equation}
J^3 |p,\sigma\rangle
  \neq\sigma\hbar\,|p,\sigma\rangle.
\end{equation}
As in the case of a massive particle,
$|p,\sigma\rangle$ cannot be
a simultaneous eigenstate of $J^3$ and $\vec{P}$
due to the motion on the $xy$-plane.
The label $\sigma$ denotes not the eigenvalue of $J^3$
but the helicity.
In fact,
since
\[
\left[\,U(R_y(\vartheta))\,U(R_z(\varphi))\,\right]
\,(\vec{J}\cdot\hat{p})\,
\left[\,U(R_y(\vartheta))\,U(R_z(\varphi))\,\right]^{-1}
\]
is equal to $J^3$ and $[J^3,K^3]=0$, we obtain
\begin{equation}
\vec{J}\cdot\hat{p}\, |p,\sigma\rangle
 =\sigma\hbar\,|p,\sigma\rangle.
\end{equation}
We normalize the state by
the invariant delta function (\ref{invdel}) with $M=0$,
\begin{equation}
\langle p',\sigma'|p,\sigma\rangle
 =\left(\frac{|\vec{p}|}{\kappa}\right)
  \,\delta^3(\vec{p'}-\vec{p})\,\delta_{\sigma'\sigma}.
\end{equation}

The state defined in Eq.~(\ref{defp0}) transforms
under the Lorentz transformation
$\Lambda\munu$ as
\begin{equation}
U(\Lambda)\, |p,\sigma\rangle
=U(L(\Lambda p))\,U(W(\Lambda,p))\, |k,\sigma\rangle^{\textrm{std}},
\end{equation}
where
\begin{equation}
W(\Lambda,p)\munu=
\left[L^{-1}(\Lambda p)\,\Lambda\,L(p)\right]\munu.
\label{defw0}
\end{equation}
By the definition (\ref{deflp0}),
this transformation does not change
the standard momentum (\ref{stdk0}),
\begin{equation}
 W(\Lambda,p)\munu k^{\nu}=k^\mu.
\end{equation}
That is, $W(\Lambda,p)\munu$ is an element of $ISO(2)$.
Therefore, from Eq.~(\ref{hel0}), we find that
\begin{equation}
 U(\Lambda)\, |p,\sigma\rangle
  = e^{i\gamma(\Lambda,p)\sigma}\,|\Lambda p,\sigma\rangle,
\label{trans0}
\end{equation}
where $\gamma(\Lambda,p)$ is given from
the decomposition (\ref{iso2}) of $W(\Lambda,p)\munu$.
Note that the helicity is Lorentz invariant
since the Lorentz transformation changes
only the phase factor of the state.

In the case of the photon,
the helicity $\sigma$ is either $+1$ or $-1$.
The states $|p,\pm1\rangle$ are
circularly polarized states.
The transformations of these states are trivial
except for the helicity-dependent phase factor.
However,
in linearly polarized bases,
\begin{equation}
 |p;\zeta\rangle_\pm
  \equiv \frac{1}{\sqrt{2}}\left[\,e^{i\zeta}|p,+1\rangle
     \pm e^{-i\zeta}|p,-1\rangle\right],
\label{linpol}
\end{equation}
the transformation law becomes
\begin{equation}
 U(\Lambda)\, |p;\zeta\rangle_\pm
  = |\Lambda p;\zeta+\gamma(\Lambda,p)\rangle_\pm.
\end{equation}
This means that the plane of polarization is rotated
by the angle $\gamma(\Lambda,p)$
due to the Lorentz transformation.

Here we explicitly calculate the transformation law
in a specific example.
Suppose that in the laboratory frame
the particle is moving along the $x$-axis
with a four-momentum,
\begin{equation}
p^\mu_\pm=(\kappa e^\xi,\pm\kappa e^\xi,0,0).
\label{p0pm}
\end{equation}
Note that we cannot obtain
the particle moving in the opposite direction by
$\xi\to-\xi$ unlike the massive particle.
This is because massless particles move
at the speed of light and thus their motion cannot be reversed.
The standard Lorentz transformation (\ref{deflp0}) becomes
\begin{equation}
L(p_+)\munu =\left[R_y^{-1}(\pi/2)B_z(\xi)\right]\munu,
\end{equation}
for $p_+$ and
\begin{equation}
L(p_-)\munu=
\left[R_z^{-1}(\pi)R_y^{-1}(\pi/2)B_z(\xi)\right]\munu,
\end{equation}
for $p_-$.
Note the range of
the angles below Eq.~(\ref{resang}).
We then introduce an observer moving
in the general direction
$(\cos\theta,\sin\theta\sin\phi,\sin\theta\cos\phi)$
with the velocity $V=c\,\tanh\chi$.
The corresponding Lorentz transformation is given by
\begin{equation}
\Lambda\munu=
\left[R_x(\phi)R_y(\theta)B_x^{-1}(\chi)
R_y^{-1}(\theta)R_x^{-1}(\phi)\right]\munu.
\end{equation}
(Note that the rotation
$[R_y^{-1}(\theta)R_x^{-1}(\phi)]\munu$
brings the direction of the observer's motion
into the $x$-axis.)
The state $|p,\sigma\rangle$ in the laboratory frame
is described by $U(\Lambda)|p,\sigma\rangle$
in the observer's rest frame.

A straightforward calculation shows that
\begin{equation}
W(\Lambda,p_\pm)\munu=
  \left(\begin{array}{cccc}
    \ast &              \ast  &             \ast & \ast   \\
    \ast &              \ast  &             \ast & \ast   \\
    \ast & -\sin\varepsilon_\pm  & \cos\varepsilon_\pm & \ast   \\
    \ast &              \ast  &             \ast & \ast
  \end{array}\right),
\end{equation}
where the $\ast$'s denote unimportant components, and
\begin{eqnarray}
\cos\varepsilon_\pm &=&\frac{D_\pm}{\sqrt{D_\pm^2+E_\pm^2}},\\
\sin\varepsilon_\pm &=&\frac{E_\pm}{\sqrt{D_\pm^2+E_\pm^2}},
\end{eqnarray}
with
\begin{eqnarray}
D_\pm &=& 1\mp\sinh\chi\cos\theta \nonumber \\
      & & +(\cosh\chi-1)(\cos^2\theta+\sin^2\theta\sin^2\phi), \\
E_\pm &=&\pm(\cosh\chi-1)\sin^2\theta\sin\phi\cos\phi.
\end{eqnarray}
Then, it is easy to see that
$\gamma(\Lambda,p_\pm)=\varepsilon_\pm$ via
the decomposition (\ref{iso2}).
The transformation law (\ref{trans0}) thus becomes
\begin{equation}
 U(\Lambda)\, |p_\pm,\sigma\rangle
  = e^{i\varepsilon_\pm\sigma}\,|\Lambda p_\pm,\sigma\rangle.
\end{equation}
In terms of the linearly polarized bases (\ref{linpol}),
this gives
\begin{equation}
 U(\Lambda)\, |p_\pm;\zeta\rangle_\pm
  = |\Lambda p_\pm;\zeta+\varepsilon_\pm\rangle_\pm.
\label{mass0}
\end{equation}
Note that $\varepsilon_\pm$ depends neither on
the standard momentum $\kappa$
nor the momentum of the particle characterized by parameter $\xi$.
On the other hand,
the rotational symmetry about the $x$-axis is
apparently broken since $\varepsilon_\pm$ depends on $\phi$.
This is because the $z$-axis is
a special axis that labels the states in the massless case.
A different choice of the $z$-axis leads to
a redefinition of the phase factor of the state.

\begin{figure}
\begin{center}
\includegraphics[scale=0.65]{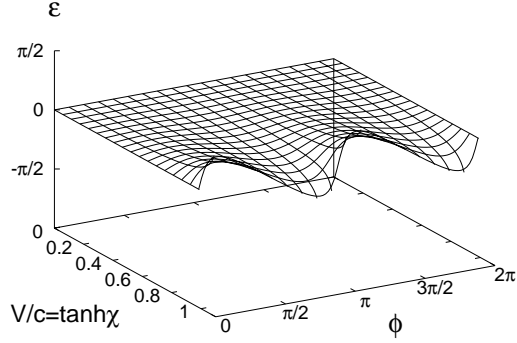}
\end{center}
\caption{\label{fig4}The rotation angle $\varepsilon$
of the plane of polarization for the orthogonal observer
as a function of $V/c=\tanh\chi$ and $\phi$.}
\end{figure}

If the observer is moving in the same direction
as the particle (i.e. $\theta=0$),
the angles $\varepsilon_\pm$ are zero.
If the observer is moving in the $z$ direction
(i.e. $\theta=\pi/2$ and $\phi=0$),
we again find that $\varepsilon_\pm$ are zero,
unlike the massive case.
This is a consequence of the fact that the state is labeled
in such a manner that the $z$-axis is a special axis.
Actually, in a situation in which
the observer is moving in the direction
$(0,\sin\phi,\cos\phi)$ on the $yz$-plane,
$\varepsilon_\pm$ take non-trivial values given by
\begin{equation}
\tan\varepsilon_\pm\equiv\pm\tan\varepsilon=
\pm\frac{(\cosh\chi-1)\sin\phi\cos\phi}{1+(\cosh\chi-1)\sin^2\phi}.
\label{defeps}
\end{equation}
The value of $\varepsilon$ as a function of $\chi$ and $\phi$
is shown in Fig.~\ref{fig4}.
If $\chi=0$ or $\phi=0$, $\pi/2$, $\pi$, $3\pi/2$,
$\varepsilon$ is zero.
When $\chi\to\infty$, we obtain
\begin{equation}
\varepsilon=
\left\{\begin{array}{cc}
   \frac{\pi}{2}-\phi & (0<\phi<\pi)   \\[5pt]
   \frac{3\pi}{2}-\phi &(\pi<\phi<2\pi)
\end{array}\right. .
\label{exteps}
\end{equation}
In this limit, there are discontinuities
at $\phi=0$, $\pi$ due to the singular behavior,
\begin{equation}
 \lim_{\phi\to0,\pi}\lim_{\chi\to\infty}\varepsilon \neq
 \lim_{\chi\to\infty}\lim_{\phi\to0,\pi}\varepsilon.
\end{equation}
For a given $\chi$, the maximum of $\varepsilon$ is attained at
\begin{equation}
\phi=\frac{1}{2}\arccos\left(\tanh^2\frac{\chi}{2}\right).
\end{equation}

In terms of the unitary operator,
we can understand the rotation of the plane
of polarization as follows:
When $\theta=\pi/2$ and $\chi$ is small,
the Lorentz transformation $\Lambda\munu$ becomes
\begin{eqnarray}
U(\Lambda) &=& e^{-\frac{i}{\hbar}
     \left(K^2\sin\phi+K^3\cos\phi\right)\chi} \nonumber \\
    &\simeq & e^{-\frac{i}{\hbar}K^3\chi\cos\phi}
              e^{-\frac{i}{\hbar}K^2\chi\sin\phi}
          e^{\frac{i}{4\hbar}J^1 \chi^2 \sin2\phi}.
\end{eqnarray}
The last factor rotates the momentum $p^\mu$
pointing in the $x$ direction
through an angle $\chi^2\sin2\phi/4$ about the $x$-axis
and gives rise to the phase factor.
The other factors contribute only to
$U(S(\alpha,\beta))$, which is irrelevant
in the case of the photon.

\section{\label{sec:epr}EPR Correlation}
In non-relativistic quantum mechanics,
the EPR correlation is often discussed
using the EPR state
(or one of the Bell states),
\begin{equation}
 |\phi\rangle=\frac{1}{\sqrt{2}}
   \Bigl[\, |\uparrow\rangle|\downarrow\rangle-
      |\downarrow\rangle|\uparrow\rangle \,\Bigr],
\label{singlet}
\end{equation}
which is the spin-singlet state of two spin-$1/2$ particles.
This state is also referred to as
an entangled state of two particles.
Moreover, this state has an important property
of the perfect anti-correlation in any direction of space.
The eigenstates of spin in the direction $\hat{n}$,
denoted by $|\uparrow\{\hat{n}\}\rangle$ and
$|\downarrow\{\hat{n}\}\rangle$,
are connected with $|\uparrow\rangle$
and $|\downarrow\rangle$ by
\begin{eqnarray}
 |\uparrow\rangle &=&
  \alpha\,|\uparrow\{\hat{n}\}\rangle+
      \beta\,|\downarrow\{\hat{n}\}\rangle, \\
 |\downarrow\rangle &=&
  \gamma\,|\uparrow\{\hat{n}\}\rangle+
      \delta\,|\downarrow\{\hat{n}\}\rangle,
\end{eqnarray}
where $\alpha$, $\beta$, $\gamma$, and $\delta$ are
the components of a unitary matrix.
Since $\alpha\delta-\beta\gamma$ is the determinant of
the unitary matrix which is equal to unity
except for an overall phase factor,
the state $|\phi\rangle$ can be written as
\begin{equation}
 |\phi\rangle=\frac{1}{\sqrt{2}}
 \Bigl[\,|\uparrow\{\hat{n}\}\rangle|\downarrow\{\hat{n}\}\rangle-
 |\downarrow\{\hat{n}\}\rangle|\uparrow\{\hat{n}\}\rangle \,\Bigr].
\end{equation}
This means that,
if the spins of both particles are measured
in the \emph{same} direction $\hat{n}$,
the results are \emph{always} anti-correlated,
irrespective of the choice of $\hat{n}$.
This property comes from the fact that
the spin-singlet state is
isotropic and has no preferred direction.

We note that other Bell states do not possess such a property.
For example, a spin-triplet state
\begin{equation}
 |\tilde{\phi}\rangle=\frac{1}{\sqrt{2}}
   \Bigl[\, |\uparrow\rangle|\uparrow\rangle +
      |\downarrow\rangle|\downarrow\rangle \,\Bigr]
\label{triplet}
\end{equation}
is also an entangled state.
If the spins of both particles are measured
in the $z$ direction,
the results are always correlated
(rather than anti-correlated).
However, in terms of the eigenstates of the $y$-component,
the state $|\tilde{\phi}\rangle$ is rewritten as
\begin{equation}
 |\tilde{\phi}\rangle=\frac{1}{\sqrt{2}}
   \Bigl[\, |\uparrow\{y\}\rangle|\downarrow\{y\}\rangle +
      |\downarrow\{y\}\rangle|\uparrow\{y\}\rangle \,\Bigr].
\label{triplet2}
\end{equation}
This means that the measurements of the $y$-component
are always anti-correlated.
Moreover,
the measurements of the spin in the direction
$\hat{m}=(0,1/\sqrt{2},1/\sqrt{2})$ are
completely random because
\begin{eqnarray}
|\tilde{\phi}\rangle
  &=&\frac{1}{2}
 \Bigl[\,i\,|\uparrow\{\hat{m}\}\rangle|\downarrow\{\hat{m}\}\rangle
  +i\,|\downarrow\{\hat{m}\}\rangle|\uparrow\{\hat{m}\}\rangle
 \nonumber \\
  & &
 +|\uparrow\{\hat{m}\}\rangle|\uparrow\{\hat{m}\}\rangle
+|\downarrow\{\hat{m}\}\rangle|\downarrow\{\hat{m}\}\rangle\,\Bigr].
\end{eqnarray}
Although the entanglement is independent of the basis,
the correlation depends on the basis for the measurement.

The spin-singlet state is
invariant under the spatial rotation.
However, it is not invariant under the Lorentz transformation.
This is because the Poincar\'{e} group $ISO(1,3)$
is larger than the spatial rotation group $SO(3)$.
Thus, in some moving frames,
the spin-singlet state is mixed
with the spin-triplet states by
the Lorentz transformation,
losing the property of anti-correlation.

\subsection{Massive Case}
\begin{figure}
\begin{center}
\includegraphics[scale=0.5]{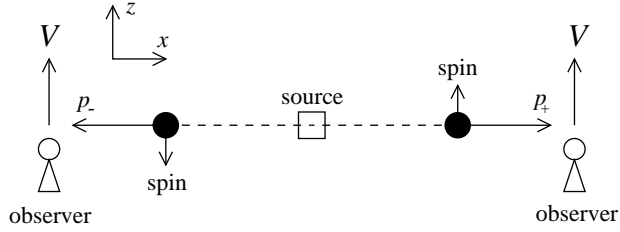}
\end{center}
\caption{\label{fig5}The relativistic EPR experiment
in the laboratory frame.}
\end{figure}

We now formulate a relativistic EPR experiment
for massive particles.
Suppose that a pair of spin-$1/2$ particles
with total spin zero are
moving away from each other along the $x$-axis,
as illustrated in Fig.~\ref{fig5}.
This system is described in the laboratory frame
by a relativistic EPR state
\begin{equation}
 |\psi\rangle=\frac{1}{\sqrt{2}}
   \Bigl[\, |p_+,\uparrow\rangle|p_-,\downarrow\rangle-
      |p_+,\downarrow\rangle|p_-,\uparrow\rangle \,\Bigr],
\label{origstate}
\end{equation}
where
\begin{equation}
  p^\mu_\pm=(Mc\,\cosh\xi,\pm Mc\,\sinh\xi,0,0).
\end{equation}
Note here that we explicitly specify
the momentum of each particle
since the Wigner rotation
depends on it as in Eq.~(\ref{trans}).
We then assume that
two observers who perform the measurements of the spins
are both moving in the $z$ direction
at the same velocity (\ref{velobs})
with respect to the laboratory frame.
(See Fig.~\ref{fig5}.)

\begin{figure}
\begin{center}
\includegraphics[scale=0.45]{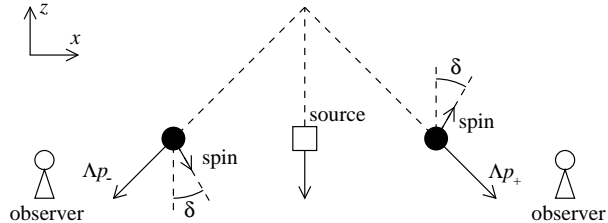}
\end{center}
\caption{\label{fig6}The relativistic EPR experiment
in the observers' frame.}
\end{figure}

In a frame in which the two observers are at rest,
the observers describe
the relativistic EPR state (\ref{origstate})
as $U(\Lambda)|\psi\rangle$.
Using the transformation formulas
(\ref{mass1}) and (\ref{mass2}),
we find that
\begin{eqnarray}
&& U(\Lambda)|\psi\rangle= \nonumber \\
&& \frac{1}{\sqrt{2}}
 \Biggl[\cos\delta\Bigl(
 |\Lambda p_+,\uparrow\rangle|\Lambda p_-,\downarrow\rangle-
 |\Lambda p_+,\downarrow\rangle|\Lambda p_-,\uparrow\rangle\Bigr)
   \nonumber \\
&& +\sin\delta\Bigl(
    |\Lambda p_+,\uparrow\rangle|\Lambda p_-,\uparrow\rangle+
    |\Lambda p_+,\downarrow\rangle|\Lambda p_-,\downarrow\rangle
  \Bigr) \Biggr],
\label{relstate}
\end{eqnarray}
where $\delta$ is given by Eq.~(\ref{defdel}).
Note that the spin of the second particle is rotated
through the angle $-\delta$
since it is moving in the negative $x$-direction.
The \emph{same} unitary transformation thus acts on
each spin in \emph{different} ways.
The situation is depicted in Fig.~\ref{fig6}.

From Eq.~(\ref{relstate}), we can see that
the spin-singlet state is mixed with the spin-triplet state
which has the same form as Eq.~(\ref{triplet}).
Thus, in the observers' frame,
even if the spins are measured in the $z$ direction,
the results are not always anti-correlated.
In the extreme case of $\xi\to\infty$ and $\chi\to\infty$,
they are perfectly correlated rather than anti-correlated.
We here emphasize that
the $z$ direction in the observers' frame is
also the $z$ direction in the laboratory frame,
since the observers are moving in the $z$ direction.
More generally, the directions that are parallel
in the laboratory frame remain parallel
in the observers' frame.
Nevertheless, the results of the spin measurements
in the same direction
are not always anti-correlated in the observers' frame,
unlike in the non-relativistic case.
(The measurements of the spin $y$-component
remain anti-correlated for any $\xi$ and $\chi$,
as Eq.~(\ref{triplet2}).)
We thus conclude that the perfect anti-correlation
in an arbitrary direction is
not maintained in all inertial frames.

One might think that
this result contradicts the nonlocality that is regarded
as an inherent property of quantum mechanics.
For example, after the first observer have measured
the spin of the first particle of an EPR pair
in the $z$ direction and obtained
a result $\uparrow$,
the state of the second particle becomes
\begin{equation}
\cos\delta \, |\Lambda p_-,\downarrow\rangle
+\sin\delta \, |\Lambda p_-,\uparrow\rangle.
\end{equation}
In this state, the spin $z$-component
does not have a definite value.
One might consider that this contradicts
the EPR argument~\cite{AhLeHw}.
However, the spin component
in the direction $(-\sin2\delta,0,\cos2\delta)$
have the definite value $\downarrow$.
The above contradiction is therefore superficial and
the perfect EPR correlation is maintained
in \emph{different} directions.
Note that, unlike the correlation that we discuss here,
the entanglement is invariant
under the Lorentz transformation
because the Wigner rotation is
a local unitary transformation~\cite{AlsMil02}.

\subsection{Massless Case}
We next formulate a relativistic EPR experiment
for massless particles with helicity $\pm1$.
Since the Lorentz transformation of
the circularly polarized states $|p,\pm1\rangle$
are trivial as in Eq.~(\ref{trans0}),
one might think that
there is no effect due to the Lorentz transformation.
However, in the linearly polarized basis (\ref{linpol}),
the plane of polarization is rotated by
the Lorentz transformation.
We thus consider a relativistic EPR state
\begin{equation}
 |\psi\rangle=\frac{1}{\sqrt{2}}
   \Bigl[\, |p_+;\zeta\rangle_+|p_-;\zeta\rangle_-
     -|p_+;\zeta\rangle_-|p_-;\zeta\rangle_+ \,\Bigr],
\label{origstate0}
\end{equation}
where $p_\pm$ is defined by Eq.~(\ref{p0pm}).
Note that the measurements of the polarization
with respect to the same angle
are always anti-correlated,
irrespective of the angle.
Both of the two observers are assumed
to move in the direction $(0,\sin\phi,\cos\phi)$
on the $yz$-plane at the velocity $V=c\,\tanh\chi$.

In the common rest frame of the observers,
the relativistic EPR state (\ref{origstate0}) is
seen as $U(\Lambda)|\psi\rangle$.
Using the transformation law~(\ref{mass0}), we find that
\begin{eqnarray}
 U(\Lambda)|\psi\rangle
 &=&\frac{1}{\sqrt{2}}
\Bigl[\,|\Lambda p_+;\zeta+\varepsilon\rangle_+
  |\Lambda p_-;\zeta-\varepsilon\rangle_- \nonumber \\
 & & -|\Lambda p_+;\zeta+\varepsilon\rangle_-
  |\Lambda p_-;\zeta-\varepsilon\rangle_+ \,\Bigr],
 \label{relstate0}
\end{eqnarray}
where $\varepsilon$ is given by Eq.~(\ref{defeps}).
Thus, the measurements of the polarization
with respect to the same angle
are not always anti-correlated in this frame.

\section{\label{sec:bell}Bell's Inequality}
Suppose that one particle is performed
a measurement either $Q$ or $R$,
and the other particle is performed
a measurement either $S$ or $T$.
Each measurement produces an outcome $+1$ or $-1$.
Then, any local realistic theory predicts that
the following Bell's inequality holds~\cite{Bell64,CHSH69,GHSZ90}:
\begin{equation}
E(QS)+E(RS)+E(RT)-E(QT)\le 2,
\label{bellin}
\end{equation}
where $E(QS)$ denotes
the expectation value of QS, etc.
However, in quantum theory, this inequality may be violated.
For example,
if we choose the observables as
\begin{eqnarray}
Q &=& \sigma_z, \nonumber \\
R &=& \sigma_y,  \nonumber \\
S &=& \frac{-\sigma_y-\sigma_z}{\sqrt{2}}, \\
T &=& \frac{-\sigma_y+\sigma_z}{\sqrt{2}},\nonumber 
\end{eqnarray}
where $\sigma_y$ and $\sigma_z$ are the Pauli matrices,
then for the spin-singlet state (\ref{singlet}),
we obtain
\begin{equation}
\langle QS\rangle+\langle RS\rangle+
\langle RT\rangle-\langle QT\rangle= 2\sqrt{2}.
\end{equation}
This is the case of the maximum violation
of Bell's inequality due to the perfect anti-correlation
of the spin-singlet state.

In the preceding section,
we have shown that
the property of the anti-correlation is not maintained
if the measurements are performed by the moving observers.
Thus, it is interesting to see its influence
on the violation of this Bell's inequality.
(By introducing moving observers,
Bell-type theorems for Lorentz-invariant realistic theories
have been discussed in Refs.~\cite{Hardy92,Perciv98}.
However, they have no use for
the Lorentz transformation of the state.)

\subsection{Massive Case}
To read out the spin component,
we first define spin operators
\begin{eqnarray}
 \sigma_x(p) &=& \frac{1}{C(p)}\Bigl[\,
   |p,\uparrow\rangle\langle p,\downarrow|+
    |p,\downarrow\rangle\langle p,\uparrow| \,\Bigr],  \\
 \sigma_y(p) &=& \frac{1}{C(p)}\Bigl[\,
   -i\,|p,\uparrow\rangle\langle p,\downarrow|+
   i\, |p,\downarrow\rangle\langle p,\uparrow| \,\Bigr],  \\
 \sigma_z(p) &=& \frac{1}{C(p)}\Bigl[\,
   |p,\uparrow\rangle\langle p,\uparrow|-
    |p,\downarrow\rangle\langle p,\downarrow| \,\Bigr],
\end{eqnarray}
where $C(p)$ is a normalization factor,
\begin{equation}
 C(p)=\langle p,\uparrow|p,\uparrow\rangle=
      \langle p,\downarrow|p,\downarrow\rangle.
\end{equation}
(Note that $C(p)$ would diverge to infinity.
However, by using a wave packet
rather than the momentum eigenstate,
the divergence can be removed, in practice.)
These operators behave as the Pauli matrices
for the states $|p,\uparrow\rangle$ and $|p,\downarrow\rangle$.
For example,
\begin{eqnarray*}
  \sigma_z(p)\,|p,\uparrow\rangle  &=& |p,\uparrow\rangle,   \\
  \sigma_z(p)\,|p,\downarrow\rangle  &=& -|p,\downarrow\rangle, \\
  \sigma_x(p)\,|p,\uparrow\rangle  &=& |p,\downarrow\rangle.
\end{eqnarray*}
We also define the expectation value of an operator $O$ as
\begin{equation}
\langle O\rangle\equiv
   \frac{\langle\psi|O|\psi\rangle}{\langle\psi|\psi\rangle}.
\end{equation}

We consider the same situation as in Fig.~\ref{fig6}
and choose the observables to be measured as
\begin{eqnarray}
Q &=& \sigma_z(\Lambda p_+),  \nonumber \\
R &=& \sigma_y(\Lambda p_+),  \nonumber \\
S &=& \frac{-\sigma_y(\Lambda p_-)-\sigma_z(\Lambda p_-)}{\sqrt{2}}, \\
T &=& \frac{-\sigma_y(\Lambda p_-)+\sigma_z(\Lambda p_-)}{\sqrt{2}}, \nonumber
\end{eqnarray}
where $Q$ and $R$ are the spin $z$- and $y$- components
of the first particle, and
$S$ and $T$ are the spin components
in the directions $(0,-1/\sqrt{2},-1/\sqrt{2})$
and $(0,-1/\sqrt{2},1/\sqrt{2})$
of the second particle.
This set of the observables gives rise to
the maximum violation of Bell's inequality
in non-relativistic quantum theory.
We, however, obtain in relativistic quantum theory,
\begin{equation}
\langle QS\rangle+\langle RS\rangle+
\langle RT\rangle-\langle QT\rangle= 2\sqrt{2}\,\cos^2\delta,
\label{massivebell}
\end{equation}
for the relativistic EPR state (\ref{relstate})
in the moving observers' frame.
If either the particle or the observer is at rest,
the relativistic EPR state
violates Bell's inequality (\ref{bellin}).
The amount of violation \emph{prima facie} decreases
as their velocities increase~\cite{Czacho97}.
In the extreme case of $\xi\to\infty$ and $\chi\to\infty$,
the right-hand side of Eq.~(\ref{massivebell}) becomes zero.

One might think that this means
a restoration of local realistic theories.
However, this conclusion is not correct
since different sets of observables or different states
still violate Bell's inequality.
For example,
if we rotate the directions of the measurements
in accordance with the Wigner rotation
by the replacement
\begin{equation}
\sigma_z(\Lambda p_+)\to
\sigma_z(\Lambda p_+)\cos\delta+\sigma_x(\Lambda p_+)\sin\delta
\end{equation}
for the first particle and by the replacement
\begin{equation}
\sigma_z(\Lambda p_-) \to
\sigma_z(\Lambda p_-)\cos\delta-\sigma_x(\Lambda p_-)\sin\delta
\end{equation}
for the second particle,
Bell's inequality turns out to be maximally violated,
\begin{equation}
\langle QS\rangle+\langle RS\rangle+
\langle RT\rangle-\langle QT\rangle= 2\sqrt{2}.
\end{equation}
Instead, if we can use the spin-singlet state
in the observers' frame (rather than the laboratory frame),
\begin{equation}
 \frac{1}{\sqrt{2}} \Bigl[\,
|\Lambda p_+,\uparrow\rangle|\Lambda p_-,\downarrow\rangle-
|\Lambda p_+,\downarrow\rangle|\Lambda p_-,\uparrow\rangle \,\Bigr],
\end{equation}
Bell's inequality is again maximally violated.
An apparent decrease in the degree of
violation of Bell's inequality results from the fact
that the Lorentz transformation rotates
the directions of the spins in a different manner;
however, since the rotations are local transformations,
they preserve the perfect anti-correlation in appropriately
chosen different direction.

\subsection{Massless Case}
We define the polarization operator
with respect to the angle $\zeta$,
\begin{eqnarray}
 P(p,\zeta)
  &=& \frac{1}{C(p)}\Bigl[\,
     e^{-2i\zeta}|p,-\rangle\langle p,+|   \nonumber \\
  & & \qquad\qquad+e^{2i\zeta} |p,+\rangle\langle p,-|\,\Bigr],
\end{eqnarray}
where $C(p)$ is again the normalization factor,
\begin{equation}
 C(p)=\langle p,+|p,+\rangle=\langle p,-|p,-\rangle.
\end{equation}
This operator reads out the polarization
with respect to the angle $\zeta$,
\begin{equation}
 P(p,\zeta)\,|p;\zeta\rangle_\pm = \pm|p;\zeta\rangle_\pm.
\end{equation}
We then choose the observables to be measured as
\begin{eqnarray}
Q &=& P\left(\Lambda p_+,0\right),  \nonumber \\
R &=& P\left(\Lambda p_+,\frac{\pi}{4}\right), \nonumber  \\
S &=& P\left(\Lambda p_-,-\frac{3\pi}{8}\right), \\
T &=& P\left(\Lambda p_-,-\frac{\pi}{8}\right),\nonumber
\end{eqnarray}
under the same situation as in the preceding section.
The maximal violation of the inequality is achieved by
this set of the observables in non-relativistic quantum theory.
However, in relativistic quantum theory,
we obtain
\begin{equation}
\langle QS\rangle+\langle RS\rangle+
\langle RT\rangle-\langle QT\rangle= 2\sqrt{2}\cos4\varepsilon,
\end{equation}
for the relativistic EPR state (\ref{relstate0})
in the observers' frame.
The right-hand side vanishes
if $\varepsilon=\pm\pi/8$ or $\pm3\pi/8$.
This is achieved for $\phi=(2n+1)\pi/8$ $(n=0,1,\cdots,7)$
in the limit of $\chi\to\infty$
as can be seen from Eq.~(\ref{exteps}).
The right-hand side can also be negative, unlike in the massive case.

Instead, if we choose the observables
in accordance with the rotation of
the plane of polarization,
\begin{eqnarray}
Q &=& P\left(\Lambda p_+,\varepsilon\right),  \nonumber \\
R &=& P\left(\Lambda p_+,\frac{\pi}{4}+\varepsilon\right),  \nonumber \\
S &=& P\left(\Lambda p_-,-\frac{3\pi}{8}-\varepsilon\right), \\
T &=& P\left(\Lambda p_-,-\frac{\pi}{8}-\varepsilon\right),\nonumber 
\end{eqnarray}
or if we use the state
\begin{equation}
 \frac{1}{\sqrt{2}}
   \Bigl[\, |\Lambda p_+;\zeta\rangle_+|\Lambda p_-;\zeta\rangle_-
     -|\Lambda p_+;\zeta\rangle_-|\Lambda p_-;\zeta\rangle_+ \,\Bigr],
\end{equation}
Bell's inequality is maximally violated.

\section{\label{sec:discuss}Summary and Discussion}
In the present paper,
we have formulated the EPR gedankenexperiment
within relativistic quantum theory.
Using the Poincar\'{e} group and its representation,
we have shown that
the perfect anti-correlation in the \emph{same} direction
deteriorates if the measurements are performed
by moving observers.
The perfect anti-correlation
in all the directions cannot be maintained
in all inertial frames.
Note that this effect is not generated by
the Galilean transformation
in non-relativistic quantum theory.
The spin-singlet state in the laboratory frame is
mixed with the spin-triplet state in the observers' frame,
because the Poincar\'{e} group $ISO(1,3)$
is larger than the spatial rotation group $SO(3)$.
Moreover, the degree of the violation of Bell's inequality
\emph{prima facie} decreases
with increasing the velocity of the observers.
To observe the perfect anti-correlation and
the maximal violation of Bell's inequality,
we have to perform the measurements in
appropriately chosen \emph{different} directions.

We now discuss a relevance to quantum information.
The entangled state are often utilized in quantum communication,
such as quantum teleportation~\cite{BBCJPW93}
and quantum cryptography~\cite{Ekert91,BeBrMe92}.
Since the entanglement is preserved under
the Lorentz transformation~\cite{AlsMil02},
the entangled state can be utilized even by moving observers.
Nevertheless, this does not mean that
the quantum communication succeeds without conditions.
We have to rotate the direction of the measurement
depending on the relative motion between
the EPR source and the observers
in order to utilize the perfect anti-correlation and
the maximal violation of Bell's inequality.
If the observers unfortunately do not know the relative motion,
the accuracy of the quantum communication is limited.
For example, suppose that one observer wants
to teleport an unknown state to the other observer
by quantum teleportation using the EPR state provided,
a posteriori, by the EPR source.
We further assume that they do not know
their velocities each other.
In non-relativistic quantum theory,
the teleportation will perfectly succeed
even in such a situation.
On the other hand, in relativistic quantum theory,
the fidelity of the teleportation is reduced
due to the interference by the Wigner rotation.
Although this interference may be small in practice,
it is interesting to recognize in principle
what kind of ignorance fundamentally limits
the accuracy of the quantum communication
in various situations.

We finally explore an experimental possibility to detect
the above effect.
While it seems not feasible to
move detectors (such as the Stern-Gerlach apparatus)
at relativistic velocities,
we may move the EPR source.
By special relativity,
moving detectors are equivalent to a moving EPR source
(see Fig.~\ref{fig6}).
It would be feasible to move the EPR source
at relativistic velocities by
the accelerator or the cosmic ray.
For example, we may utilize a spinless particle accelerated
to a relativistic velocity by the accelerator,
which decays into a pair of spin-$1/2$ particles
moving in opposite directions
in the rest frame of the original particle.
The velocity $V$ corresponds to that of the original particle
in the laboratory frame and
the velocity $v$ corresponds to that of the resultant particles
in the original particle's frame,
caused by the decay process.
Using rest detectors,
we could see a reduction in the anti-correlation
in the same direction.
The locations of the detectors are decided by the angle of
the momentum rotation (\ref{momang}).
While there are many technological challenges,
we should be able to observe the relativistic effect by
this kind of experiment.

\section*{Acknowledgments}
H.T. was partially supported
by JSPS Research Fellowships for Young Scientists.
This work was supported by a Grant-in-Aid for
Scientific Research (Grant No. 11216204)
by the Ministry of Education, Science, Sports, and Culture
of Japan, by the Toray Science Foundation,
and by the Yamada Science Foundation.

\appendix
\section*{Appendix}

\section{Poincar\'{e} Group}
We summarize here the Poincar\'{e} group
as it serves as a basic mathematical tool.
The Poincar\'{e} group is
the symmetry of the Minkowski spacetime
which is a set of
coordinate transformations
that leave the world length,
\begin{equation}
ds^2=-c^2dt^2+dx^2+dy^2+dz^2,
\end{equation}
invariant.
This set includes translations, rotations,
and boosts (or Lorentz transformations).
For notational simplicity,
we adopt the following conventions:
\begin{itemize}
\item The spacetime coordinates $(ct,x,y,z)$ are denoted
as $(x^0,x^1,x^2,x^3)$.
\item Greek letters $\mu,\nu,\cdots$ represent
the spacetime indices $0,1,2,3$.
\item Latin letters $i,j,k,\cdots$ represent
the spatial indices $1,2,3$.
\item 3-dimensional vectors $(a^1,a^2,a^3)$ are
denoted as $\vec{a}$.
\item 4-dimensional contravariant vectors $(a^0,a^1,a^2,a^3)$ are
denoted as $a^\mu$.
\item 4-dimensional covariant vectors $(-a^0,a^1,a^2,a^3)$ are
denoted as $a_\mu$.
\item Repeated indices are to be summed.
\end{itemize}

The temporal and spatial translations are
written together as
\begin{equation}
 x^\mu \to x'^\mu=x^\mu+a^\mu,
\end{equation}
where $a^\mu$ is a constant vector.

The spatial rotation belongs to the $SO(3)$ subgroup
of the Poincar\'{e} group.
The rotation about a unit vector $\hat{n}$
through an angle $\theta$ is written as
\begin{equation}
 x^\mu \to x'^\mu=R_{\hat{n}}(\theta)\munu x^\nu,
\end{equation}
where $R_{\hat{n}}(\theta)\munu$ is an orthogonal matrix
that acts non-trivially only on the spatial coordinates.
For example, the rotations about the $z$-axis and
about the $y$-axis are described by
\begin{equation}
R_z(\theta)\munu=
  \left(\begin{array}{cccc}
    1 &          0  &          0 & 0   \\
    0 & \cos\theta  & \sin\theta & 0   \\
    0 &-\sin\theta  & \cos\theta & 0   \\
    0 &          0  &          0 & 1
  \end{array}\right)
\end{equation}
and
\begin{equation}
R_y(\theta)=
  \left(\begin{array}{cccc}
    1 &          0  & 0 &  0            \\
    0 & \cos\theta  & 0 & -\sin\theta   \\
    0 &          0  & 1 &  0            \\
    0 & \sin\theta  & 0 & \cos\theta
  \end{array}\right),
\end{equation}
respectively.

The boost is the Lorentz transformation
in a narrow sense.
The boost in the direction of a unit vector $\hat{n}$
with the rapidity $\xi$ is written as
\begin{equation}
 x^\mu \to x'^\mu=B_{\hat{n}}(\xi)\munu x^\nu,
\end{equation}
where $B_{\hat{n}}(\xi)\munu$ 
is a pseudo-orthogonal matrix,
in the sense that
\begin{equation}
\left[B_{\hat{n}}(\xi)\right]^{\mathrm{T}}
\eta B_{\hat{n}}(\xi)=\eta, \qquad
 \eta=\left(\begin{array}{cccc}
    -1 & 0 & 0 & 0 \\
     0 & 1 & 0 & 0 \\
     0 & 0 & 1 & 0 \\
     0 & 0 & 0 & 1
  \end{array}\right),
\end{equation}
and acts on both temporal and spatial components.
For example, the boost along the $x$-axis is described by
\begin{equation}
B_x(\xi)\munu=
  \left(\begin{array}{cccc}
    \cosh\xi &  \sinh\xi & 0 & 0 \\
    \sinh\xi &  \cosh\xi & 0 & 0 \\
           0 &         0 & 1 & 0 \\
           0 &         0 & 0 & 1
  \end{array}\right).
\label{boostx}
\end{equation}
Note that, if the rapidity $\xi$ is set as
\begin{equation}
\tanh\xi=-\frac{v}{c},
\end{equation}
$B_x(\xi)\munu$ in Eq.~(\ref{boostx})
reduces to the well-known formula for
the Lorentz transformation,
\begin{equation}
 \left(\begin{array}{cccc}
 \frac{1}{\sqrt{1-(v/c)^2}}  &-\frac{v/c}{\sqrt{1-(v/c)^2}}&0&0\\
-\frac{v/c}{\sqrt{1-(v/c)^2}}& \frac{1}{\sqrt{1-(v/c)^2}}  &0&0\\
  0 & 0 & 1 & 0 \\
  0 & 0 & 0 & 1
  \end{array}\right).
\end{equation}
Similarly, the boost along the $z$-axis is described by
\begin{equation}
B_z(\xi)\munu=
  \left(\begin{array}{cccc}
    \cosh\xi &  0 & 0 & \sinh\xi \\
           0 &  1 & 0 &        0 \\
           0 &  0 & 1 &        0 \\
    \sinh\xi &  0 & 0 & \cosh\xi
  \end{array}\right).
\end{equation}
The set of all boosts does not constitutes
a subgroup of the Poincar\'{e} group
because it is not closed.

The Lorentz transformation in a wider sense,
which we shall denote as $\Lambda\munu$,
describes a combination of rotation and boost.
It consists of all the matrices that satisfy
the pseudo-orthogonal condition,
\begin{equation}
\Lambda^{\mathrm{T}}
\eta\Lambda=\eta.
\end{equation}
The set of all such transformations constitutes
a subgroup of the Poincar\'{e} group,
known as the Lorentz group $SO(1,3)$.

In quantum theory, the symmetry transformation
is represented by a unitary operator which acts
on the Hilbert space.
Since relativistic quantum theory means quantum theory with
the Poincar\'{e} symmetry,
we have unitary operators corresponding to
all the Poincar\'{e} transformations.

The unitary operator corresponding to
the translation $a^\mu=(a^0,\vec{a})$ is given by
\begin{equation}
 U(a)=\exp\left[\frac{i}{\hbar}
   \left(\frac{Ha^0}{c}-\vec{P}\cdot\vec{a}\right)\right],
\label{translation}
\end{equation}
where $H$ and $\vec{P}$ are
the Hamiltonian and momentum operators
which are combined to form a four-momentum operator,
$\mathcal{P}^\mu=(H/c,\vec{P})$.
Equation (\ref{translation}) can thus be written as
\begin{equation}
 U(a)=\exp\left(-\frac{i}{\hbar}
      \mathcal{P}^\mu a_\mu\right).
\end{equation}
For a particle with mass $M$ and
velocity $\vec{v}=d\vec{x}/dt$,
the four-momentum is defined by
\begin{equation}
 p^\mu=\left(\frac{Mc}{\sqrt{1-|\vec{v}/c|^2}},
     \frac{M\vec{v}}{\sqrt{1-|\vec{v}/c|^2}}\right).
\end{equation}
Since this four-momentum satisfies the basic relation
in special relativity,
\begin{equation}
 p_\mu p^\mu=-(p^0)^2+|\vec{p}|^2=-M^2c^2,
\end{equation}
the Hamiltonian and momentum operators must satisfy
\begin{equation}
 \mathcal{P}_\mu \mathcal{P}^\mu=
  -\frac{H^2}{c^2}+|\vec{P}|^2=-M^2c^2.
\label{oprelation}
\end{equation}
On the other hand,
for a massless particle with momentum $\vec{p}$
(and energy $E=|\vec{p}|c$),
the four-momentum is defined by
\begin{equation}
 p^\mu=\left(|\vec{p}|,\vec{p}\right).
\end{equation}
Thus, the Hamiltonian and momentum operators satisfy
Eq.~(\ref{oprelation}) with $M=0$.

The unitary operators corresponding to
the rotation $R_{\hat{n}}(\theta)\munu$ and
the boost $B_{\hat{n}}(\xi)\munu$ are given by
\begin{eqnarray}
 U(R_{\hat{n}}(\theta)) &=&
  \exp\left(\frac{i}{\hbar}\vec{J}\cdot\hat{n}\theta\right), 
   \label{defJ}\\
 U(B_{\hat{n}}(\xi)) &=&
  \exp\left(\frac{i}{\hbar}\vec{K}\cdot\hat{n}\xi\right),
   \label{defK}
\end{eqnarray}
where $\vec{J}$ is the total angular momentum operator
and $\vec{K}$ is the boost operator.
(In classical mechanics, these correspond to
$x^ip^j-x^jp^i$ and $x^ip^0-x^0p^i$, respectively.)
The operators $\vec{J}$ and $\vec{K}$ are combined to
form an antisymmetric tensor of rank 2,
$\mathcal{J}^{\mu\nu}$ $(=-\mathcal{J}^{\nu\mu})$, where
\begin{eqnarray*}
(\mathcal{J}^{23},\mathcal{J}^{31},\mathcal{J}^{12})
 &=& \vec{J}, \\
(\mathcal{J}^{10},\mathcal{J}^{20},\mathcal{J}^{30})
 &=& \vec{K}.
\end{eqnarray*}
Since the explicit forms of these operators
are unnecessary for our present purpose,
we simply view Eqs.~(\ref{defJ}) and (\ref{defK}) as
the definitions of $\vec{J}$ and $\vec{K}$.

In order for these unitary operators
to represent the Poincar\'{e} group,
the generators $H$, $\vec{P}$, $\vec{J}$, and $\vec{K}$ must
satisfy the commutation relations~\cite{Weinbe95},
\begin{eqnarray}
  \left[J^i, J^j\right] &=& i\hbar\epsilon_{ijk}J^k, \nonumber \\
  \left[J^i, K^j\right] &=& i\hbar\epsilon_{ijk}K^k, \nonumber \\
  \left[K^i, K^j\right] &=& -i\hbar\epsilon_{ijk}J^k,\nonumber \\
  \left[J^i, P^j\right] &=& i\hbar\epsilon_{ijk}P^k, \nonumber \\
  \left[K^i, P^j\right] &=& \frac{i\hbar}{c}H\delta_{ij},      \\
  \left[K^i, H\right]   &=& i\hbar cP^i,                 \nonumber \\
  \left[J^i, H\right]   &=& 0,                   \nonumber \\
  \left[P^i, H\right]   &=& 0.                   \nonumber
\end{eqnarray}
When $\vec{K}=0$,
we obtain the usual commutation relations of $ISO(3)$
which consists of spatial translations and rotations.


\begin{thebibliography}{10}

\bibitem{EiPoRo35}
A. Einstein, B. Podolsky, and N. Rosen, Phys. Rev. {\bf 47},  777  (1935).

\bibitem{Bohm89}
D. Bohm,  in {\em Quantum Theory} (Dover, New York, 1989), pp.\ 611--623.

\bibitem{Bloch67}
I. Bloch, Phys. Rev. {\bf 156},  1377  (1967).

\bibitem{HelKra70}
K.-E. Hellwig and K. Kraus, Phys. Rev. D {\bf 1},  566  (1970).

\bibitem{AhaAlb}
Y. Aharonov and D.~Z. Albert, Phys. Rev. D {\bf 24},  359  (1981);
Phys. Rev. D {\bf 29},  228  (1984).

\bibitem{HelKon83}
T.~M. Helliwell and D.~A. Konkowski, Am. J. Phys. {\bf 51},  996  (1983).

\bibitem{GhGrPe90}
G.~C. Ghirardi, R. Grassi, and P. Pearle, Found. Phys. {\bf 20},  1271  (1990).

\bibitem{SuaSca}
A. Suarez and V. Scarani, Phys. Lett. A {\bf 232},  9  (1997);
A. Suarez, Phys. Lett. A {\bf 236},  383  (1997).

\bibitem{ZBGT01}
H. Zbinden, J. Brendel, N. Gisin, and W. Tittel, Phys. Rev. A {\bf 63},  022111
   (2001).

\bibitem{Ohnuki88}
Y. Ohnuki, {\em Unitary Representations of the Poincar{\'e} group and
  Relativistic Wave Equations} (World Scientific, Singapore, 1988).

\bibitem{Weinbe95}
S. Weinberg, {\em The Quantum Theory of Fields} (Cambridge University Press,
  Cambridge, 1995), chap. 2.5.

\bibitem{Wigner39}
E.~P. Wigner, Ann. Math. {\bf 40},  149  (1939).

\bibitem{PeScTe02}
A. Peres, P.~F. Scudo, and D.~R. Terno, Phys. Rev. Lett. {\bf 88},  230402
  (2002).

\bibitem{AlsMil02}
P.~M. Alsing and G.~J. Milburn, Quantum Information and Computation {\bf 2},
  487  (2002).

\bibitem{TerUed02}
H. Terashima and M. Ueda, quant-ph/0204138.

\bibitem{GinAda02}
R.~M. Gingrich and C. Adami, quant-ph/0205179.

\bibitem{AhLeHw}
D. Ahn, H.-j. Lee, and S.~W. Hwang, quant-ph/0207018;
D. Ahn, H.-j. Lee, Y.~H. Moon, and S.~W. Hwang, quant-ph/0209164.

\bibitem{Bell64}
J.~S. Bell, Physics {\bf 1},  195  (1964).

\bibitem{CHSH69}
J.~F. Clauser, M.~A. Horne, A. Shimony, and R.~A. Holt, Phys. Rev. Lett. {\bf
  23},  880  (1969).

\bibitem{GHSZ90}
D.~M. Greenberger, M.~A. Horne, A. Shimony, and A. Zeilinger, Am. J. Phys. {\bf
  58},  1131  (1990).

\bibitem{Hardy92}
L. Hardy, Phys. Rev. Lett. {\bf 68},  2981  (1992).

\bibitem{Perciv98}
I.~C. Percival, Phys. Lett. A {\bf 244},  495  (1998).

\bibitem{Czacho97}
In a situation where the two particles are co-moving,
Bell's inequality has been discussed
using a different definition of the relativistic spin
by M. Czachor, Phys. Rev. A {\bf 55},  72  (1997).
He obtained a decrease in the degree of violation
due to the motion of the particle rather than the observer.
Contrary to this result,
Bell's inequality is unaffected in our formulation
if the two particles are co-moving.
Czachor's effect is thus different from ours.

\bibitem{BBCJPW93}
C.~H. Bennett {\it et~al.}, Phys. Rev. Lett. {\bf 70},  1895  (1993).

\bibitem{Ekert91}
A.~K. Ekert, Phys. Rev. Lett. {\bf 67},  661  (1991).

\bibitem{BeBrMe92}
C.~H. Bennett, G. Brassard, and N.~D. Mermin, Phys. Rev. Lett. {\bf 68},  557
  (1992).

\end{thebibliography}
\end{document}